\documentclass[12pt]{article}
\usepackage{a41}

\usepackage{palatino}
\usepackage{xcolor}
\usepackage{amsmath}
\usepackage{booktabs}
\usepackage{microtype}

\usepackage{cite}
\usepackage{amsmath}
\usepackage{amssymb}
\usepackage{color}

\newcommand{\mathgraphics}[2][]{\vcenter{\hbox{\includegraphics[#1]{#2}}}}

\newcommand{\mpl}{\Lambda}
\DeclareMathOperator{\diag}{diag}

\usepackage[rflt]{floatflt}
\usepackage{float}
\usepackage{slashed}

\def\b0{\beta_0}

\usepackage{array}

\usepackage[english]{babel}
\usepackage[latin1]{inputenc}
\usepackage[T1]{fontenc}
\usepackage{ae}

\usepackage{url}

\usepackage{amsmath, amsthm, amssymb}
\newtheorem{thm}{Theorem}[section]

\newtheorem{definition}[thm]{Definition}

\newcommand{\ep}{\varepsilon}

\usepackage{rotating}

\usepackage{graphicx}

\newcounter{mmacnt}
\def\restartmma{\setcounter{mmacnt}{0}}
\restartmma \catcode`|=\active
\def|#1|{\mathrm{#1}}
\catcode`|=12
\newenvironment{mma}{
 \par\smallskip
 \catcode`|=\active
 \parskip=0pt\parindent=0pt 
 \small
 \def\In##1\\{%
\def\linebreak{\hfill\break\null\qquad}%
\refstepcounter{mmacnt}
\hangindent=2.5em\hangafter=0
\leavevmode
\llap{\tiny\sffamily n[\arabic{mmacnt}]:=\kern.5em}%
\mathversion{bold}\footnotesize$\displaystyle##1$\normalsize
\mathversion{normal}\par
 }%
 \def\Print##1\\{%
\def\linebreak{\hfill\break}%
\hangindent=2.5em\hangafter=0
\leavevmode ##1\par}%
 \def\Out##1\\{%
\def\linebreak{$\hfill\break\null\hfill$}%
\kern\abovedisplayskip\par
\hangindent=2.5em\hangafter=0
\leavevmode
\llap{\tiny\sffamily Out[\arabic{mmacnt}]=\kern.5em}
\footnotesize$\displaystyle##1$\normalsize\hfill\null\par
\kern\belowdisplayskip
 }%
 \def\Warning##1##2\\{%
\def\linebreak{\hfill\break}%
\hangindent=2.5em\hangafter=0
\leavevmode
{\scriptsize##1 : ##2}\par}%
}{%
 \par\smallskip
}

\usepackage{color}

\newenvironment{fshaded}{%
\MakeFramed {\FrameRestore}
}%
{\endMakeFramed}

\def\b0{\beta_0}

\def\Gp0{{\Gamma^{'}_0}}
\def\Gp1{{\Gamma^{'}_1}}
\def\Gp2{{\Gamma^{'}_2}}

\allowdisplaybreaks[4]

\begin{document}
\setlength{\baselineskip}{0.515cm}

\sloppy
\thispagestyle{empty}
\begin{flushleft}
DESY 19--029
\hfill {\tt arXiv:1902.11180[gr-qc]}
\\
DO--TH 19/01\\
\end{flushleft}

\mbox{}
\vspace*{\fill}
\begin{center}

{\Large\bf Five-Loop Static Contribution to the}

\vspace*{2mm}
{\Large\bf Gravitational Interaction Potential}

\vspace*{2mm}
{\Large\bf of Two Point Masses}

\vspace{3cm} \large
{\large J.~Bl\"umlein, A.~Maier, and P.~Marquard}

\vspace{1.cm}
\normalsize
{\it   Deutsches Elektronen--Synchrotron, DESY,}\\
{\it   Platanenallee 6, D--15738 Zeuthen, Germany}


\end{center}
\normalsize
\vspace{\fill}
\begin{abstract}
\noindent
We compute the static contribution to the gravitational interaction potential of two point masses
in the velocity-independent five-loop (and 5th post-Newtonian) approximation to the harmonic coordinates
effective action in a direct calculation. The computation is performed using effective field methods based
on Feynman diagrams in momentum-space in $d = 3 - 2\ep$ space dimensions. We also reproduce the previous
results including the 4th post-Newtonian order.
\end{abstract}

\vspace*{\fill}
\noindent
\newpage

\section{Introduction}
\label{sec:1}

\vspace*{1mm}
\noindent
The interpretation of the signals detected in gravitational wave interferometers like LIGO and VIRGO \cite{LIGO}
requires a very accurate knowledge of the binary dynamics of large coalescing masses. This also applies to the
planned projects like INDIGO, LISA Pathfinder and the Einstein telescope \cite{PLANNED}. The increasing improvements
of the detector sensitivity requires highly precise theoretical predictions.

Different approaches are used as the
effective one-body formalism \cite{BD99,BD00,Damour:2014afa,Damour:2016gwp,Damour:18,VINES}, numerical relativity
\cite{PRET,CAMP,BAK}, the self-force formalism \cite{MINO,QUINN}, the post-Newtonian (PN)
\cite{EINSTEIN1915,DROSTE,Einstein:1938yz,FOCK,PLEB,CHAND,Kimura:1972ew,OHTA1,OHTA2,Damour:1985mt,WS93,Blanchet:1987wq,
Jaranowski:1997ky,DJS1,BF1,deAndrade:2000gf,DJS2,DamourPoincare,BjerrumBohr:2002kt,Kol:2007bc,Gilmore:2008gq,
Kol:2010ze,Kol:2010si,Foffa:2011ub,Foffa:2012rn,Bini:2013zaa,DJS3,JS1,Bernard:2015njp,
Damour:2016abl,Bernard:2016wrg,Foffa:2016rgu,Damour:2017ced,Bini:2017wfr,
BjerrumGrav,Plefka:2018dpa,KOSOWER,Antonelli:2019ytb}
and post-Minkowskian approach \cite{Berotti,Kerr,BP1,Porti1,WG1,Porti2,Bel1,WEST1,Foffa:2013gja,
Damour:2016gwp,Damour:18,Cheung,BjerrumGrav,Bern:2019nnu} and effective field theory methods
\cite{Damour:1995kt,EFT2}; for surveys see~Refs.~\cite{Futamase:2007zz,BLANCHET14,PORTO16,SJ18,BK18,LEVI18}.

In this letter we calculate a first contribution to the fifth post-Newtonian approximation: the five-loop {\it static}
gravitational interaction potential between two non-spinning point masses. The two-particle force receives a series of
different higher order corrections, which can be parameterized by a small parameter $\epsilon$, cf. Ref.~\cite{Futamase:2007zz},
\begin{equation}
F = F_N
+ \sum_{k=1}^\infty \epsilon^{2k} F_{kPN}
+ \epsilon^4 F_{SO}
+ \epsilon^4 F_{QO}
+ \epsilon^5 F_{RR}
+ \epsilon^6 F_{1PN SO}
+ \epsilon^6 F_{1PN QO}
+ \epsilon^6 F_{OO}
+ \epsilon^6 F_{SS}
+ \epsilon^6 F_{TO} + ...
\end{equation}
Here $F_N$ denotes Newton's force \cite{NEWTON}, $F_{kPN}$ is the $k$th post-Newtonian force, $F_{RR}$ the 2.5 PN
radiation reaction force, $F_{SO}$ and $F_{1PN SO}$ etc. are the spin-orbit coupling force and their post-Newtonian
corrections, $F_{QO}$ and $F_{1PN QO}$ etc. are the quadrupole-orbit coupling force and its post-Newtonian
corrections. $F_{OO}, F_{SS}$ and $F_{TO}$ denote the octupole-orbit coupling force, the spin-spin coupling force,
and tidal-orbit coupling force, respectively.\footnote{Radiation and spin effects are discussed in
Refs.~\cite{OHTA2,Schaefer:1986rd,Jaranowski:1996nv,Luna,GR1,GPT1,GR2,Chester,GLP1,LP1,Shen}
and
Refs.~\cite{S1,S2,S3,Holstein:2008sx,Levi:2011eq,
S5,S6,S7,S8,S10,S11,Vaidya,Levi:2015ixa,Levi:2015msa,Guevara1,Guevara2,Chung1},
respectively.
Radiative multipoles have been computed in various earlier work, reviewed in e.g. Refs.
\cite{Futamase:2007zz,
BLANCHET14,
PORTO16,
SJ18,
BK18,
LEVI18}.}
We will concentrate here on the post-Newtonian corrections of the attraction of two spinless masses in the
following, using non-relativistic gravitational fields obtained by a temporal Kaluza-Klein reduction \cite{KK} followed
by a Weyl rescaling \cite{WEYL}. The corresponding
action has been derived in Ref.~\cite{Kol:2010si}. In the representation time derivatives of arbitrary order occur, which
also introduce higher derivatives of the accelerations $a_{1(2)}$. Since the Lagrange density of General Relativity is
of second order, these terms shall be eliminated by adding suitable double (multiple)-zero terms \cite{Damour:1985mt}.
One aims on the terms of the order
\begin{equation}
\label{eq:ORD}
\sim  \frac{G_N^k}{r^k} m_1^{l} m_2^{k+1-l},~~~l \in [1,k],
\end{equation}
where $(k-1)$ labels the  $k$th post-Newtonian approximation. Here $G_N$ denotes Newton's constant, $r =
|\mathbf{r}|$ the distance of the two masses and $m_{1(2)}$ are the two point masses.

We calculate the contribution to the fifth post-Newtonian approximation in the static limit, i.e. leaving the
velocity-dependent contributions for a later work. The virial theorem \cite{CLAUSIUS} relates  $m_1
\mathbf{v}_1^2 + m_2 \mathbf{v}_2^2 \sim \frac{G_N}{r} m_1 m_2$ on temporal average, where $\mathbf{v}_{1(2)}$
denote the velocities of the two point masses. Therefore velocity terms have to be considered at this order in general.
This also applies to higher derivatives of the velocities, which can finally be mapped
to terms $\propto (G_N^k/r^k) v_i^l$ by applying the equation of motion.

We first outline the basic formalism and present then the details of the calculation. Finally, we compare to the
results in the literature including the fourth post-Newtonian approximation.
\section{Basic Formalism}
\label{sec:2}

\vspace*{1mm}
\noindent
The action of General Relativity for the present problem consists of the following three
components,
\begin{equation}
  \label{eq:S_GR}
  S_{GR} = S_\text{pp} + S_\text{EH} + S_\text{GF}\,,
\end{equation}
where $S_\text{pp}, S_\text{EH}$ and $S_\text{GF}$ are the point-particle-, the Einstein-Hilbert- ,
\cite{Blanchet:2003gy}, and the gauge fixing contributions. Following \cite{Kol:2010si} we parameterize the
Riemann metric by
\begin{equation}
  \label{eq:KK_metric}
  g_{\mu\nu} = e^{2\phi/\mpl}
  \begin{pmatrix}
    -1 & A_j/\mpl\\
    A_i/\mpl & e^{-c_d\phi/\mpl}\gamma_{ij}-A_i A_j/\mpl^2
  \end{pmatrix}\,.
\end{equation}
Here, the scalar field $\phi$, 3-vector field $A_i$ and tensor field $\gamma_{ij}$ are introduced
which parameterize the ten components of the metric tensor and $\Lambda$ and $c_d$ are given by
\begin{eqnarray}
\Lambda^{-1} &=& \sqrt{32 \pi G_N},
\\
c_d  &=& 2\frac{d-1}{d-2},
\end{eqnarray}
where $d := 3 - 2\ep$ and $G_N \rightarrow G_N (\mu)^{3-d}$ in $d$ dimensions, with $\mu$ a mass scale. In the
Newtonian limit of flat space--time we adopt the `mostly plus' convention
$g_{\mu\nu} \to \eta_{\mu\nu} = \diag(-1,+1,+1,+1)$, with $\eta_{\mu\nu}$ the Minkowskian metric.
The contribution to the spatial metric $\gamma_{ij}$ is parameterized by
\begin{eqnarray}
\gamma_{ij} &=& \delta_{ij} + \sigma_{ij}/\mpl,
\end{eqnarray}
where $\delta_{ij}$ denotes Kronecker's symbol. The gravitational field amplitude is given by
\begin{equation}
h^{\alpha\beta} = \sqrt{-g} g^{\alpha\beta} - \eta^{\alpha\beta}
\end{equation}
and the harmonic coordinate condition reads
\begin{equation}
 \partial_\mu h^{\alpha\mu} = 0.
\end{equation}

The point-particle action
\begin{eqnarray}
  \label{eq:S_pp}
S_\text{pp} &=& - \sum_{k=1}^2 m_k \int d\tau_k = - \sum_{k=1}^2 m_k \int dt
\sqrt{-g_{\mu\nu} \frac{dx_k^\mu}{dt} \frac{dx_k^\nu}{dt}} \nonumber\\
&=& - \sum_{k=1}^2 m_k \int dt e^{\phi/\Lambda} \sqrt{(1- \mathbf{A}.\mathbf{v}_k)^2
- e^{-c_d \phi/\Lambda} \gamma_{ij} v_k^i v_k^j}
\end{eqnarray}
describes the point masses themselves.

The dynamics of the metric in one temporal and $d$ spatial dimensions are
captured by the Einstein-Hilbert action, \cite{Kol:2010si},
\begin{equation}
  \label{eq:S_EH}
S_\text{EH} = \frac{1}{16 \pi G_N} \int d^{d}x~dt \sqrt{-g} R[g]
= S^{(0)}(\gamma, A, \phi)
+ S^{(1)}(\gamma, A, \phi)
+ S^{(2)}(\gamma, A, \phi),
\end{equation}
sorting w.r.t. the occurring number of time-derivatives. For the static contribution calculated in the present paper
only
$S^{(0)}$ is relevant.
\begin{eqnarray}
S^{(0)}(\gamma, A, \phi) &=& - \frac{1}{16 \pi G_N} \int d^d x~dt \sqrt{\gamma} \Biggl[ - R[\gamma] +
\frac{1}{2} c_d
\gamma^{ij} \partial_i \phi \partial_j \phi
- \frac{1}{4} e^{c_d\phi} \gamma^{ik} \gamma^{jl} F_{ij} F_{kl}\Biggr],
\end{eqnarray}
where  $R[\gamma]$ denotes the Ricci scalar and $F_{ij} = \partial_i A_j - \partial_i A_j$. The
determinant $\gamma$ may be represented by
\begin{eqnarray}
\gamma  &=& 1 + \sum_{k=1}^\infty \frac{\hat{\sigma}_k}{\Lambda^k} =
1  + \frac{1}{\Lambda} \text{tr}(\sigma)
   + \frac{1}{2 \Lambda^2} \left[\text{tr}^2(\sigma) - \text{tr}\left(\sigma^2\right)\right]
   + \frac{1}{6 \Lambda^3} \left[\text{tr}^3(\sigma) + 2 \text{tr}\left(\sigma^3\right)
\right. \nonumber\\ && \left.
- 3
\text{tr}\left(\sigma^2\right)
\text{tr}(\sigma)\right] + O\left(\frac{1}{\Lambda^4}\right).
\end{eqnarray}
Finally, $S_\text{GF}$ is a gauge
fixing action. We use the harmonic gauge
\begin{eqnarray}
\label{eq:S_GF}
S_\text{GF} &=& -\frac{1}{32 \pi G_N} \int d^d x~dt \sqrt{-g}\Gamma_\mu\Gamma^\nu,
\end{eqnarray}
where $\Gamma^\mu = g^{\rho\sigma} \Gamma^\mu_{\rho\sigma}$ and $\Gamma^\mu_{\rho\sigma}$ denotes the
Christoffel symbol.

The Feynman rules are derived from the path integral for the action (\ref{eq:S_GR}) expanding in $1/\Lambda$
to the desired order and retaining the terms contributing in the static limit. {This
approach has been followed before also in \cite{Gilmore:2008gq,Foffa:2011ub,Foffa:2016rgu}.}
Including the fifth post-Newtonian approximation the following relations are relevant.
The Feynman rules for the
propagators read
\begin{align}
  \label{eq:Feynman_rules_propagators}
  \phi:\quad\mathgraphics{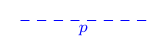}={}& - \frac{i}{2c_d\mathbf{p}^2},\\
  \sigma:\quad\mathgraphics{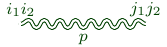}={}& -
  \frac{i}{2\mathbf{p}^2}\big(\delta_{i_1j_1}\delta_{i_2j_2}+\delta_{i_1j_2}\delta_{i_2j_1}
  + (2-c_d)\delta_{i_1i_2}\delta_{j_1j_2}\big).
\end{align}
Note that the kinetic terms are not canonically normalized,
hence the unusual form of the scalar propagator. The point-mass propagator in position space is $D(x,y) =
\Theta(x^0-y^0) \delta(\mathbf{x}-\mathbf{y})$. We can always arrange our calculation
in such a way that only symmetric combinations $D(x,y) + D(y,x) = \delta(\mathbf{x}-\mathbf{y})$ appear. In
the static case, the Fourier transform to momentum space and the resulting momentum space propagator are given by
\begin{equation}
  \label{eq:Feynman_rule_pointmass}
  \mathgraphics{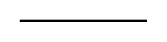} = 1\,.
\end{equation}
For the vertex Feynman rules, we choose all momenta as incoming. Choosing all momenta as outgoing or as left-to-right
will give identical expressions. In contrast to \cite{Foffa:2016rgu} we follow the usual normalization for vertices
involving multiple identical fields. For example, the coupling of a point mass to $n$ scalars does not involve a
factor of $1/n!$.
{For more than one tensor field we obtain different Feynman rules than given in \cite{Foffa:2016rgu}. However, we
agree with the result at 4PN.}
\begin{align}
  \label{eq:Feynman_rule_v_pps}
  \mathgraphics{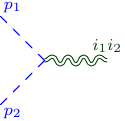} ={}&i\frac{c_d}{2\mpl}(V_{\phi\phi\sigma}^{i_1 i_2} +
V_{\phi\phi\sigma}^{i_2 i_1}) \,,\displaybreak[0]\\
  \label{eq:v_pps}
  V_{\phi\phi\sigma}^{i_1 i_2} ={}&
\mathbf{p}_1\cdot \mathbf{p}_2 \delta^{i_1i_2}- 2 p_1^{i_1} p_2^{i_2}\,,\displaybreak[0]\\
  \label{eq:Feynman_rule_v_ppss}
 \mathgraphics{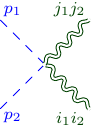} ={}&
i\frac{c_d}{16\mpl^2}(V_{\phi\phi\sigma\sigma}^{i_1i_2,j_1j_2} +
V_{\phi\phi\sigma\sigma}^{i_2i_1,j_1j_2} +
V_{\phi\phi\sigma\sigma}^{i_1i_2,j_2j_1} +
V_{\phi\phi\sigma\sigma}^{i_2i_1,j_2j_1})\,,\displaybreak[0]\\
  \label{eq:v_ppss}
V_{\phi\phi\sigma\sigma}^{i_1i_2,j_1j_2} =& \mathbf{p}_1\cdot \mathbf{p}_2
(\delta^{i_1i_2}\delta^{j_1j_2} - 2\delta^{i_1j_1}\delta^{i_2j_2})
-2(p_1^{i_1}p_2^{i_2}\delta^{j_1j_2} + p_1^{j_1}p_2^{j_2}\delta^{i_1i_2})
+ 8\delta^{i_1j_1} p_1^{i_2}p_2^{j_2}\,,\displaybreak[0]\\
  \label{eq:Feynman_rule_v_sss}
  \mathgraphics{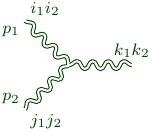} ={}&\frac{i}{32\mpl}(\tilde{V}_{\sigma\sigma\sigma}^{i_1i_2,j_1j_2,k_1k_2} +
       \tilde{V}_{\sigma\sigma\sigma}^{i_2i_1,j_1j_2,k_1k_2})\,,\displaybreak[0]\\
  \tilde{V}_{\sigma\sigma\sigma}^{i_1i_2,j_1j_2,k_1k_2} ={}&
       V_{\sigma\sigma\sigma}^{i_1i_2,j_1j_2,k_1k_2} +
       V_{\sigma\sigma\sigma}^{i_1i_2,j_2j_1,k_1k_2} +
       V_{\sigma\sigma\sigma}^{i_1i_2,j_1j_2,k_2k_1} +
  V_{\sigma\sigma\sigma}^{i_1i_2,j_2j_1,k_2k_1}
\\
V_{\sigma\sigma\sigma}^{i_1i_2,j_1j_2,k_1k_2}
={}&(\mathbf{p}_1^2+\mathbf{p}_1\cdot\mathbf{p}_2+\mathbf{p}_2^2)\*\Big(-\delta^{j_1j_2}\*\big(2\*\delta^{i_1k_1}\*
\delta^{i_2k_2}-\delta^{i_1i_2}\*\delta^{k_1k_2}\big)
\notag\\
&\quad+2\*\big[\delta^{i_1j_1}\*\big(4\*\delta^{i_2k_1}\*\delta^{j_2k_2}-\delta^{i_2j_2}\*\delta^{k_1k_2}\big)
-\delta^{i_1i_2}\*\delta^{j_1k_1}\*\delta^{j_2k_2}\big] \Big)
\notag\\
&+2\*\Big\{4\*\big(p_1^{k_2}\*p_2^{i_2}-p_1^{i_2}\*p_2^{k_2}\big)\*\delta^{i_1j_1}\*\delta^{j_2k_1}
\notag\\
&\quad+2\*\big[\big(p_1^{i_1}+p_2^{i_1}\big)\*p_2^{i_2}\*\delta^{j_1k_1}\*\delta^{j_2k_2}-p_1^{k_1}\*p_2^{k_2}\*
\delta^{i_1j_1}\*\delta^{i_2j_2}\big]
\notag\\
&\quad+\delta^{j_1j_2}\*\big[p_1^{k_1}\*p_2^{k_2}\*\delta^{i_1i_2}+2\*\big(p_1^{k_2}\*p_2^{i_2}-p_1^{i_2}\*
p_2^{k_2}\big)\*\delta^{i_1k_1}-\big(p_1^{i_1}+p_2^{i_1}\big)\*p_2^{i_2}\*\delta^{k_1k_2}\big]
\notag\\
&\quad+p_2^{j_2}\*\Big(4\*p_1^{i_2}\*\delta^{i_1k_1}\*\delta^{j_1k_2}+p_1^{j_1}\*\big(2\*\delta^{i_1k_1}\*
\delta^{i_2k_2}-\delta^{i_1i_2}\*\delta^{k_1k_2}\big)
\notag\\
&\qquad+2\*\big[\delta^{i_1j_1}\*\big(p_1^{i_2}\*\delta^{k_1k_2}-2\*p_1^{k_2}\*\delta^{i_2k_1}\big)-p_1^{k_2}\*
\delta^{i_1i_2}\*\delta^{j_1k_1}\big] \Big)
\notag\\
&\quad+p_1^{j_2}\*\Big(p_1^{j_1}\*\big(2\*\delta^{i_1k_1}\*\delta^{i_2k_2}-\delta^{i_1i_2}\*\delta^{k_1k_2}\big)
-4\*p_2^{i_2}\*\delta^{i_1k_1}\*\delta^{j_1k_2}
\notag\\
&\qquad+2\*\big[p_2^{k_2}\*\delta^{i_1i_2}\*\delta^{j_1k_1}+\delta^{i_1j_1}\*\big(2\*p_2^{k_2}\*\delta^{i_2k_1}
-p_2^{i_2}\*\delta^{k_1k_2}\big)\big]
\Big)\Big\},
  \label{eq:v_sss}
\displaybreak[0]
\\
  \label{eq:Feynman_rule_v_wwnp}
\mathgraphics{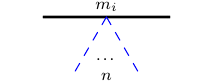} ={}& -i\frac{m_i}{\mpl^n}\,,
\end{align}
see also Ref.~\cite{Damour:2017ced}.
Here $m_i,~i=1,2$ denote the mass of the interacting world line.
\section{Feynman diagrams contributing to the potential}
\label{sec:3}

\begin{figure}[th]
  \centering
  \hskip-0.8cm
  \includegraphics[width=.15\linewidth]{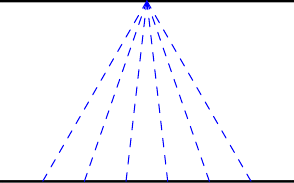}  \hspace*{2mm}
  \includegraphics[width=.15\linewidth]{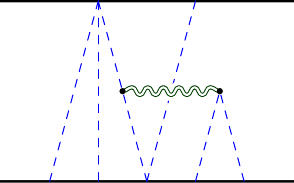} \hspace*{2mm}
  \includegraphics[width=.15\linewidth]{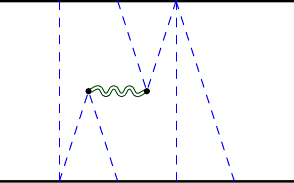} \hspace*{2mm}
  \includegraphics[width=.15\linewidth]{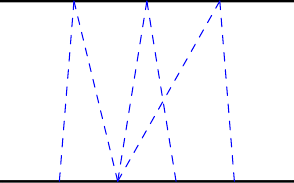} \hspace*{2mm}
  \includegraphics[width=.15\linewidth]{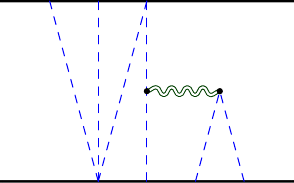}

\vspace*{3mm}
\noindent
\hspace*{-9mm}
  \includegraphics[width=.15\linewidth]{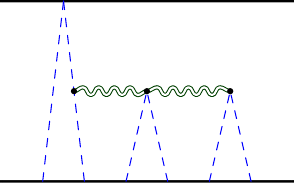}  \hspace*{2mm}
  \includegraphics[width=.15\linewidth]{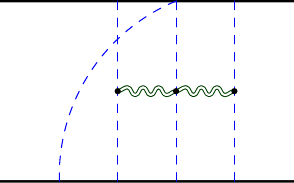} \hspace*{2mm}
  \includegraphics[width=.15\linewidth]{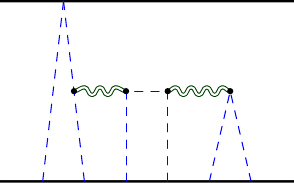} \hspace*{2mm}
  \includegraphics[width=.15\linewidth]{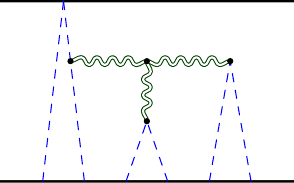} \hspace*{2mm}
  \includegraphics[width=.15\linewidth]{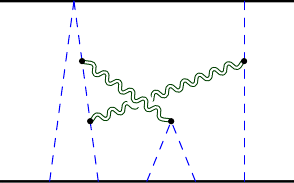}
  \caption{\small\sf Sample diagrams for the $2 \rightarrow 2$ scattering process in the static limit
  at 5PN order.}
  \label{fig:DIA1}
\end{figure}

\vspace*{1mm}
\noindent
The calculation is performed making intense use of computer algebra, the usual approach in higher order calculations
in Quantum Field Theories.\footnote{There are also computer algebraic implementations like {\tt EFTofPNG}
\cite{Levi:2017kzq} used up to 3PN in the point mass sector.}
We generate the Feynman diagrams contributing to the static interaction potential of two point masses
using {\tt QGRAF} \cite{Nogueira:1991ex}, working in momentum space and setting velocity-contributions
to zero ab initio.\footnote{For related potential calculations in QCD, see
e.g.~\cite{Smirnov:2008pn,Anzai:2009tm}.} To minimise the number
of diagrams which will not contribute to the final result of the present calculation we do
not generate self-energy diagrams, demand that the number of world line (WL) propagators
are equal to the number of loops and eliminate diagrams in which two vertices are connected
by more than one propagator. This results in the number of diagrams given in column 2
of Table~\ref{TAB1}. A series of sample diagrams is shown in Figure~\ref{fig:DIA1}.
Diagrams that factorise when cutting an arbitrary number of worldlines
correspond to multiple potential interactions and therefore yield no
additional information. We discard such diagrams (column
3). Furthermore, the contributing graphs should not have loops out of
world lines (column 4) and not contain massless tadpoles in the static
limit (column 5).
\begin{figure}[th]
  \centering
  \hskip-0.8cm
  \includegraphics[width=.15\linewidth]{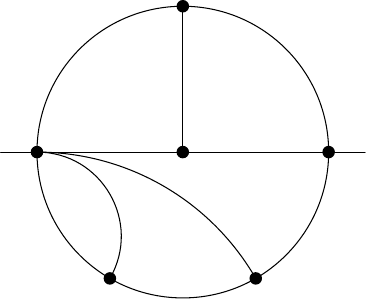}  \hspace*{2mm}
  \includegraphics[width=.15\linewidth]{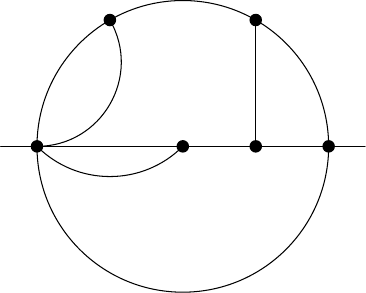} \hspace*{2mm}
  \includegraphics[width=.15\linewidth]{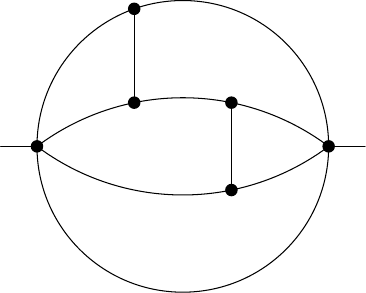} \hspace*{2mm}
  \includegraphics[width=.15\linewidth]{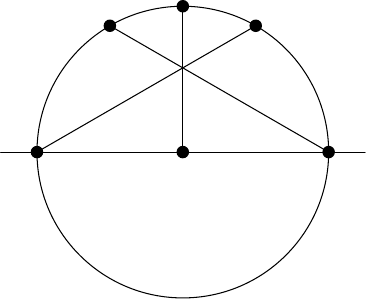} \hspace*{2mm}
  \includegraphics[width=.15\linewidth]{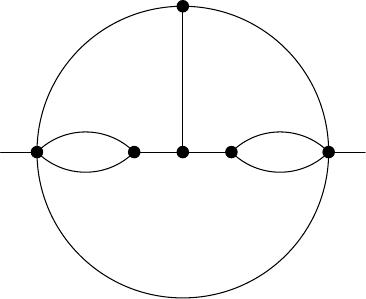}

\vspace*{3mm}
\noindent
\hspace*{-9mm}
  \includegraphics[width=.15\linewidth]{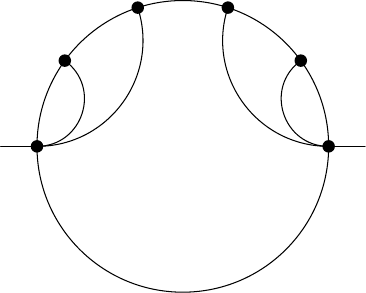}  \hspace*{2mm}
  \includegraphics[width=.15\linewidth]{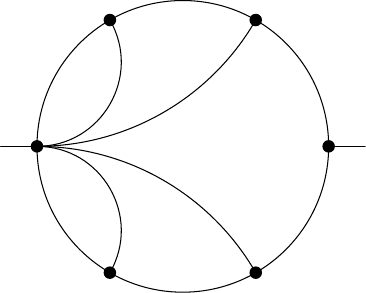} \hspace*{2mm}
  \includegraphics[width=.15\linewidth]{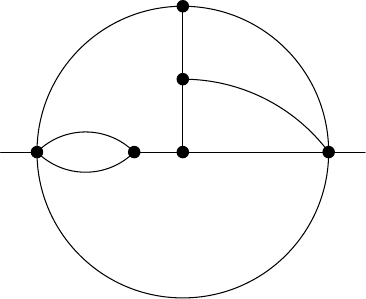} \hspace*{2mm}
  \includegraphics[width=.15\linewidth]{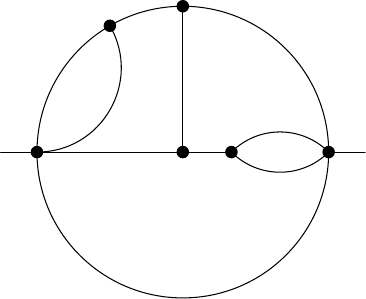} \hspace*{2mm}
  \includegraphics[width=.15\linewidth]{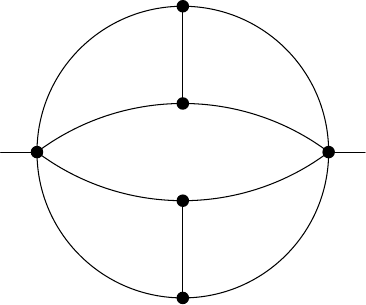}

\vspace*{3mm}
\noindent
\hspace*{-9mm}
  \includegraphics[width=.15\linewidth]{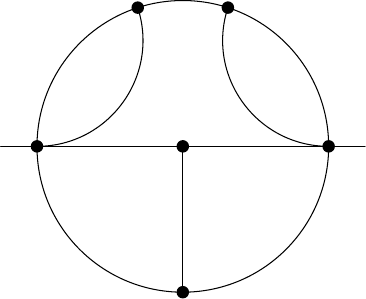}  \hspace*{2mm}
  \includegraphics[width=.15\linewidth]{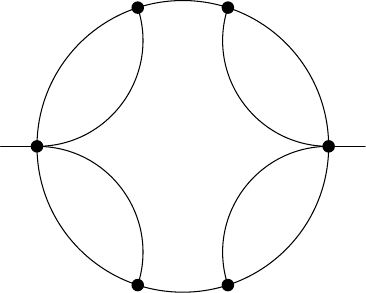} \hspace*{2mm}
  \includegraphics[width=.15\linewidth]{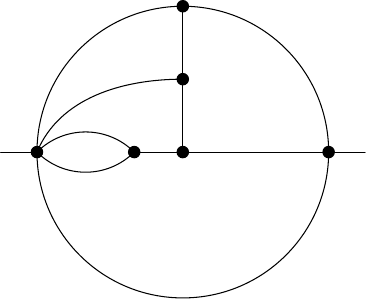} \hspace*{2mm}
  \includegraphics[width=.15\linewidth]{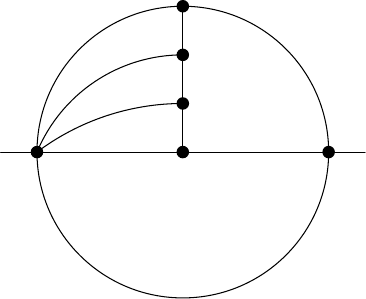} \hspace*{2mm}
  \includegraphics[width=.15\linewidth]{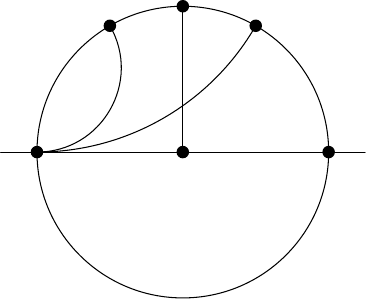}

\vspace*{3mm}
\noindent
\hspace*{-9mm}
  \includegraphics[width=.15\linewidth]{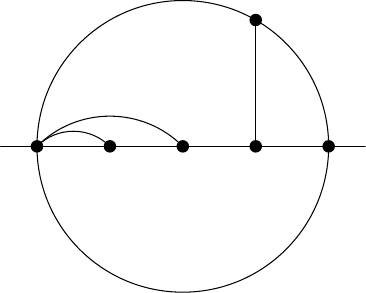}  \hspace*{2mm}
  \includegraphics[width=.15\linewidth]{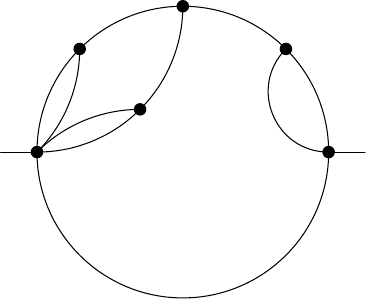} \hspace*{2mm}
  \includegraphics[width=.15\linewidth]{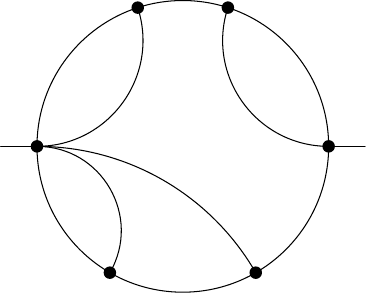} \hspace*{2mm}
  \includegraphics[width=.15\linewidth]{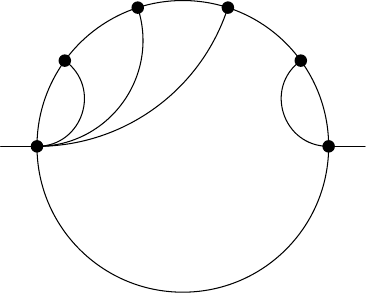} \hspace*{2mm}
  \includegraphics[width=.15\linewidth]{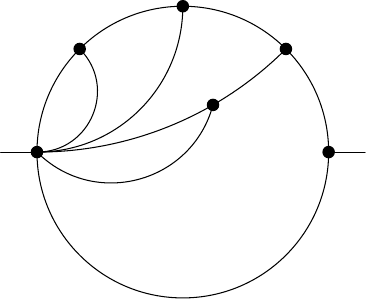}

\vspace*{3mm}
\noindent
\hspace*{-9mm}
  \includegraphics[width=.15\linewidth]{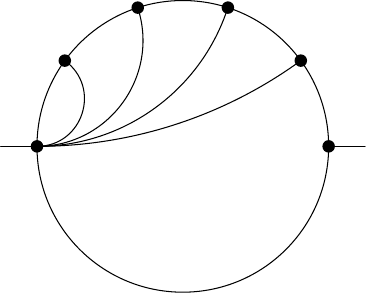}  \hspace*{2mm}
  \includegraphics[width=.15\linewidth]{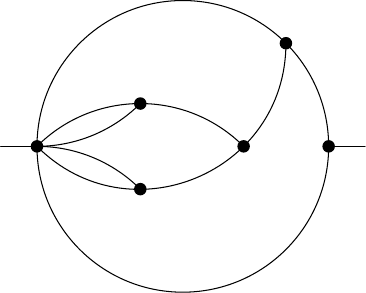}
  \caption{\small\sf The topologies contributing to the potential in the static limit at 5PN order.}
  \label{fig:DIA2}
\end{figure}
At 5PN order 27582 diagrams remain. In principle, these can be reduced to a
smaller set of diagrams by using symmetry relations under the exchange
of the world lines and all vertices on each world line, introducing
additional symmetry factors. This procedure would lead to the number of
diagrams shown in column 6. Up to 4PN order, we have checked that those
diagrams agree with those listed in~\cite{Foffa:2012rn,Foffa:2016rgu}.
However, we find that this last symmetrisation provides no benefit
in our setup and do not apply it in our calculation.

In the static limit the diagrams for $2 \to 2$ scattering can be
transformed into massless propagator-type diagrams representing the
spacial potential between the two sources. At five-loop order we
identify 22 top-level topologies, see Figure~\ref{fig:DIA2}. We then
insert the Feynman rules and perform algebraic simplifications using
{\tt FORM} \cite{Vermaseren:2000nd, Tentyukov:2007mu}. The remaining
integrals are reduced using integration by parts \cite{IBP} implemented
in the package {\tt Crusher} \cite{CRUSHER} leading to eight master
integrals, out of which four contribute. In the same way we have
calculated all the lower orders including 4PN. In the last column of
Table~\ref{TAB1} we list the number of master integrals, which turn out
to be remarkably small. The present problem is by far simpler than the
calculation of the five-loop $\beta$-function in QCD \cite{BETA5} and
massive three-loop calculations in QCD, cf.~\cite{MASS3}.
\begin{table}[H]\centering
\renewcommand{\arraystretch}{1.3}
\setlength{\tabcolsep}{11pt}
\begin{tabular}{rrrrrr@{\quad}r}
\toprule
\multicolumn{1}{c}{ }            &
\multicolumn{1}{c}{QGRAF}         &
\multicolumn{1}{c}{non fact.}     &
\multicolumn{1}{c}{no WL loops}     &
\multicolumn{1}{c}{no tadpoles}     &
\multicolumn{1}{c}{\# Diag. \cite{Foffa:2012rn}}  &
\multicolumn{1}{c}{\# MI} \\
\midrule
    N  &      1  &      1 &      1 &     1 &  1 &  -    \\
  1PN  &      2  &      2 &      2 &     2 &  1 &  1    \\
  2PN  &     19  &     19 &     19 &    15 &  5 &  1    \\
  3PN  &    360  &    276 &    258 &   122 &  8 &  1(1) \\
  4PN  &  10081  &   5407 &   4685 &  1815 & 50 &  6(1) \\
  5PN  & 332020  & 128080 & 101570 & 27582 & 154&  4(4) \\
\bottomrule
\end{tabular}
\caption[]{\small \sf Numbers of contributing diagrams at the different (post)-Newtonian levels
and master integrals. The numbers in brackets denote the number of master integrals which occur
during the reduction but do not contribute to the potential. In the  next-to-last column the number of
diagrams of equal value are given according to \cite{Foffa:2012rn}.}
\label{TAB1}
\renewcommand{\arraystretch}{1.0}
\end{table}

We have first calculated the static corrections up to 4PN in the way described above. The massless master integrals needed
are known up to three loop order from Ref.~\cite{Chetyrkin:1980pr} and at four loop order
\cite{Baikov:2010hf,Foffa:2016rgu,Damour:2017ced}. The master integrals depend on (multiple) zeta values
\cite{Blumlein:2009cf}, including $\ln(2)$. They have been compared to the numerical results given by {\tt FIESTA}
\cite{Binoth:2003ak,Smirnov:2008py,Smirnov:2009pb,Smirnov:2015mct}. Here it is useful to use the Monte Carlo
integrator {\tt Divonne} \cite{DIVONNE} of the {\tt CUBA} package \cite{Hahn:2004fe} besides {\tt VEGAS}
\cite{Lepage:1977sw}.

\begin{figure}[th]
  \centering
  \hskip-0.8cm
  \includegraphics[width=.2\linewidth]{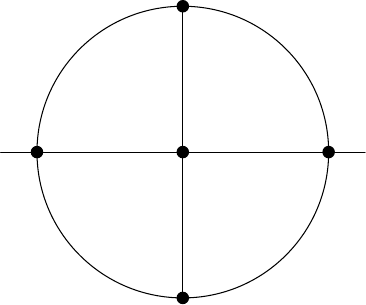}
  \caption{\small\sf The master integral $M_{36}$ contributing to the 4th post-Newtonian approximation.}
  \label{fig:MAST0}
\end{figure}
\begin{figure}[th]
  \centering
  \hskip-0.8cm
  \begin{tabular}{cccc}
  \includegraphics[width=.2\linewidth]{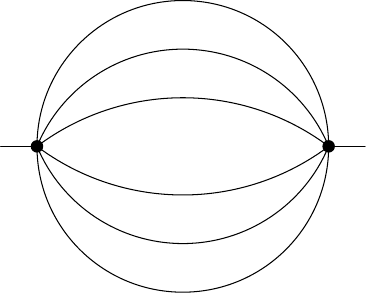} &
  \includegraphics[width=.2\linewidth]{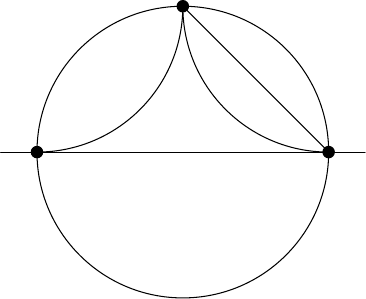} &
  \includegraphics[width=.2\linewidth]{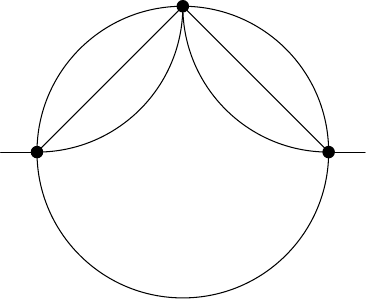} &
  \textcolor{blue}{\includegraphics[width=.2\linewidth]{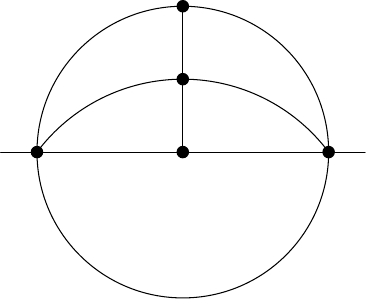}}\\[.5em]
    $M_{61}$ & $M_{72}$ & $M_{74}$ & $M_{91}$
  \end{tabular}
  \caption{\small\sf The five-loop master integrals contributing to the
    5th post-Newtonian approximation in the static limit.}
  \label{fig:MAST1}
\end{figure}
We use the $\overline{\sf MS}$-prescription in $d$ dimensions. The
one-loop two-point function is therefore defined as
\begin{equation}
\mathgraphics[width=45pt]{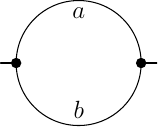} = \int \frac{d^d k}{\pi^{d/2}} \frac{1}{((p-k)^2)^a} \frac{1}{(k^2)^b}
= \frac{1}{(p^2)^{a+b-d/2}} \frac{
\Gamma\left(\frac{d}{2}-a\right)
\Gamma\left(\frac{d}{2}-b\right)
\Gamma\left(a+b-\frac{d}{2}\right)}{\Gamma(a) \Gamma(b) \Gamma\left(d-a-b\right)}\,.
\end{equation}
The contributing five loop master integrals can all be traced back to
lower loop structures by identifying effective propagator
insertions. This leads to relations of the form
\begin{equation}
M(p^2) = \int \frac{d^d k}{\pi^{d/2}} M_1\big(k^2\big) M_2\big((p-k)^2\big)\,,
\end{equation}
where $M_1, M_2$ are suitably chosen propagator insertions and the
argument of $M, M_1, M_2$ denotes the respective external momentum squared. As a
shorthand notation we define $M := M(1)$ and $S_\ep  = \exp(-\gamma_E \ep)$.

The most complicated insertion is
\begin{eqnarray}
M_{36} &=&
2 \pi^2 S_\ep^4 \Biggl[\frac{1}{\ep^2} + \frac{2}{\ep} - 2(16 -  \zeta_2) + 16 \Biggl[9 - 6 \zeta_2
\left(\frac{13}{8} -
\ln(2) \right)
- \frac{77}{6} \zeta_3\Biggr] \Biggr]\ep + O(\ep^2),
\end{eqnarray}
cf.~\cite{Foffa:2016rgu,Damour:2017ced}. A closed form representation for general values of $d$ of $M_{36}$
is not known. Here $\zeta_k = \sum_{l=1}^\infty (1/l^k),~~k \in \mathbb{N},~~k \geq 2$, denote Riemann's
$\zeta$-function at integers. With this, diagram $M_{91}$ can be
obtained as a bubble insertion of diagram $M_{36}$ into the two-point
function using
\begin{eqnarray} M_{91}(p^2)=\int \frac{d^d k}{\pi^{d/2}}
\frac{1}{((p-k)^2)^a} \frac{M_{36}}{(k^2)^b} = \frac{1}{(p^2)^{a+b-d/2}}
\frac{ \Gamma\left(\frac{d}{2}-a\right) \Gamma\left(\frac{d}{2}-b\right)
\Gamma\left(a+b-\frac{d}{2}\right)}{\Gamma(a) \Gamma(b)
\Gamma\left(d-a-b\right)} M_{36},
\end{eqnarray}
where $a = 1$ and $b=2 + 4\ep$.

The results for all master integrals contributing to the potential read\footnote{Note that typographical errors in the formula for $M_{72}$ present in earlier versions of this manuscript have been corrected.}
\begin{eqnarray}
M_{61} &=&  \frac{\Gamma\left(6 - \frac{5 d}{2}\right) \Gamma^6\left(-1 +
\frac{d}{2}\right)}{\Gamma(-6 + 3 d)}
\nonumber\\
&=& S_\ep^5 \pi^{7/2} \Biggl[\frac{2}{3}
+ \left(\frac{134}{9} + \frac{4}{3} \ln(2) \right) \ep +
   \left(\frac{5894}{27} + \frac{268}{9} \ln(2) + \frac{4}{3} \ln^2(2) + 19 \zeta_2\right) \ep^2 \Biggr]+ O(\ep^3)\,,
\nonumber\\
\\
M_{72} &=&
\frac{\Gamma \left(
        7-\frac{5 d}{2}\right) \Gamma \left(3-d\right) \Gamma\left(
        2-\frac{d}{2}\right) \Gamma^7 \left(
        -1+\frac{d}{2}\right) \Gamma(2d-5)}{\Gamma \left(5-\frac{3}{2} d\right) \Gamma(-2+d) \Gamma \left(
        -3 + \frac{3}{2} d\right) \Gamma \left(-7+3 d\right)}
\nonumber\\ &=&
- S_\ep^5 \pi^{7/2} \Biggl[
\frac{2}{\ep} +
4\left(11 +  \ln(2)\right)
+ \left(656 + 88 \ln(2) + 4 \ln^2(2) + 25 \zeta_2\right) \ep
\nonumber\\ &&
+  \left(8288 + 1312 \ln(2) + 88 \ln^2(2)
+ 550 \zeta_2
+ \frac{8}{3} \ln^3(2)
+ 50 \ln(2) \zeta_2  - \frac{2002}{3} \zeta_3\right) \ep^2
\Biggr]
\nonumber\\
&& + O(\ep^3)\,,
\\
M_{74} &=&
\frac{\Gamma \left(
        7-\frac{5 d}{2}\right) \Gamma^2 (3-d) \Gamma^7 \left(
        -1+\frac{d}{2}\right) \Gamma \left(
        -6+\frac{5 d}{2}\right)}{\Gamma \left(6-2 d\right) \Gamma^2 \left(
        -3+\frac{3 d}{2}\right) \Gamma \left(-7+3 d\right)}
\nonumber\\
&=& -S_\ep^5 \pi^{7/2} \Biggl[
\frac{4}{\ep}
+72 +8 \ln(2)
+\left(864
+144 \ln(2)
+8 \ln^2(2)
+146 \zeta_2
\right) \ep
\nonumber\\ &&
+ \left(
8640
+1728 \ln(2)
+144 \ln^2(2)
+2628 \zeta_2
+\frac{16}{3} \ln^3(2)
+292 \zeta_2 \ln(2)
-\frac{1988}{3} \zeta_3
\right) \ep^2 \Biggr]
\nonumber\\
&& + O(\ep^3)\,,
\\
M_{91} &=& 6 \pi^{7/2} S_\ep^5 \Biggl[\frac{2}{\ep} - 4(1 - \ln(2)) - \left(48 + 8 \ln(2) - 4
\ln^2(2) - 105 \zeta_2\right)
\ep
+ \Biggl(480 - 96 \ln(2)
\nonumber\\ &&
- 8 \ln^2(2)
+ \frac{8}{3} \ln^3(2) - 530 \zeta_2 + 402 \ln(2) \zeta_2 - \frac{1522}{3}
\zeta_3\Biggr) \ep^2 \Biggr] + O(\ep^3)\,.
\end{eqnarray}
\section{Results}
\label{sec:4}

\vspace*{1mm}
\noindent
Inserting the master integrals into the expression for the static contribution to the Lagrangian it turns out that
all pole contributions in the dimensional parameter $\ep$ cancel.
\begin{eqnarray}
\label{eq:Lstat}
{\cal L}^S =
{\cal L}_N^S +
{\cal L}_{PN_1}^S +
{\cal L}_{PN_2}^S +
{\cal L}_{PN_3}^S +
{\cal L}_{PN_4}^S +
{\cal L}_{PN_5}^S
\end{eqnarray}
up to the fifth post-Newtonian order. One obtains the following contributions to the static
Lagrangian
up to the 5th PN order:
\begin{align}
{\cal L}_N^S      &= \frac{G_N}{r} m_1 m_2, &\text{\cite{NEWTON}}
\\
{\cal L}_{PN_1}^S &=  -\frac{G_N^2}{2 r^2} m_1 m_2 (m_1+m_2), &\text{\cite{Einstein:1938yz}}
\\
{\cal L}_{PN_2}^S &=  \frac{G_N^3}{r^3} m_1 m_2 \left(\frac{1}{2}(m_1^2 + m_2^2) + 3 m_1
m_2\right),
&\text{\cite{Gilmore:2008gq}}
\\
{\cal L}_{PN_3}^S &=  - \frac{G_N^4}{r^4} m_1 m_2 \left[
\frac{3}{8}\left(m_1^3+m_2^3\right)
+ 6 m_1 m_2 (m_1+m_2)
\right],
&\text{\cite{Foffa:2011ub}}
\\
{\cal L}_{PN_4}^S &= \frac{G_N^5}{r^5} m_1 m_2
\left[
  \frac{3}{8} \left(m_1^4 + m_2^4\right)
+ \frac{31}{3} m_1 m_2 \left(m_1^2 + m_2^2\right)
+ \frac{141}{4} m_1^2 m_2^2 \right], &\text{\cite{Foffa:2016rgu}}
\\
  \label{eq:result_PN5}
{\cal L}_{PN_5}^S &= - \frac{G_N^6}{r^6} m_1 m_2 \Biggl[
                      \frac{5}{16} (m_1^5+m_2^5)
                    + \frac{91}{6} m_1 m_2 (m_1^3+m_2^3)
                    + \frac{653}{6} m_1^2 m_2^2 (m_1 + m_2)
\Biggr],
\end{align}
with $r = |{\bf r}_1 - {\bf r}_2|$. We agree with the above results in the literature up to ${\cal L}_{PN_4}^{S}$.
Moreover, including 5PN, all the contributions due to multiple zeta values also cancel and only
rational coefficients
remain.

To be explicit, the pole part of ${\cal L}_{PN_5}^S$ in terms of the
master integrals is given by
\begin{equation}
  \Big[{\cal L}_{PN_5}^S\Big]_{\ep^{-1}} = \frac{G_N^6}{r^6} (m_1
  m_2)^3(m_1+m_2) \pi^{-7/2}\frac{45}{32}\Big[2M_{72} -
  M_{74}\Big]_{\ep^{-1}} = 0\,,
\end{equation}
where $[X]_{\ep^{n}}$ denotes the coefficient of $\ep^n$ in $X$. With
this in mind, the finite part can be written as
\begin{equation}
  \begin{split}
    \Big[{\cal L}_{PN_5}^S\Big]_{\ep^{0}} ={}& - \frac{G_N^6}{r^6} (m_1
    m_2) \pi^{-7/2}\left\{ \frac{15}{32}
      (m_1^5+m_2^5)\Big[M_{61}\Big]_{\ep^0} + \frac{91}{4} m_1 m_2
      (m_1^3+m_2^3) \Big[M_{61}\Big]_{\ep^0} \right.\\
      &\quad+ m_1^2 m_2^2 (m_1 + m_2)
      \left( \Big[\frac{293}{4} M_{61} - \frac{45}{16}
          M_{72} +\frac{45}{32} M_{74}\Big]_{\ep^0} \right.\\
          &\left.\left.\qquad+ \Big[\frac{519}{16} M_{72} - \frac{627}{32}
        M_{74}+2 M_{91} \Big]_{\ep^{-1}}\right) \right\} \,.
  \end{split}
\end{equation}
{We remark that up to the term ${\cal L}_{PN_1}$ the results also agree to the zero--velocity limit
of \cite{Bernard:2015njp,DJS3}. In the limit $m_1 \ll m_2$ the coefficients of the leading terms
of $O(m_1(G_N m_2/r)^k)$ for $k \in \mathbb{N},~k \geq 1$ are obtained as the expansion
coefficients of the generating function $-m_1 [\sqrt{(1-x)/(1+x)}|_{x = G_N m_2/r}-1]$ in accordance with
the Schwarzschild solution of a test particle in the field of a second mass.}
\section{Conclusions}
\label{sec:conclusions}

\vspace*{1mm}
\noindent
We have calculated the five-loop correction to the gravitational
interaction potential between two static point masses. This constitutes
an important part of the effective gravitational Lagrangian at fifth
post-Newtonian order. We also agree with the corresponding results in
the lower post-Newtonian orders. The calculation of the velocity-dependent terms at
this order is work in progress.

\noindent

\subsection*{Acknowledgments}


\noindent
We would like to thank Prof. Th.~Damour and Prof.~G.~Sch\"afer for very helpful comments and P.~Nogueira for a
discussion. This project has received funding from the European Union's Horizon 2020 research and innovation
programme under the Marie Sk\/{l}odowska-Curie grant agreement No. 764850, SAGEX, and COST action CA16201: Unraveling
new physics at the LHC through the precision frontier.\\[.5em]

\noindent\textbf{Note added.}\\
A few hours before the present submission the preprint~\cite{Foffa:2019hrb} appeared, presenting an independent
calculation of the same quantity. The final results, Eq.~\eqref{eq:result_PN5} in this work and Eq.~(32)
in~\cite{Foffa:2019hrb}, are in agreement.

{\small

}
\end{document}